# X-BAND CRAB CAVITIES FOR THE CLIC BEAM DELIVERY SYSTEM

G. Burt, P.K. Ambattu, A.C. Dexter, T. Abram, Cockcroft Institute, Lancaster University, Lancaster, LA1 4YR, UK
V. Dolgashev, S. Tantawi, SLAC, Menlo Park, CA
R.M. Jones, Cockcroft Institute, Manchester University, Manchester, M13 9PL, UK

*Abstract*

The CLIC machine incorporates a 20 mrad crossing angle at the IP to aid the extraction of spent beams. In order to recover the luminosity lost through the crossing angle a crab cavity is proposed to rotate the bunches prior to collision. The crab cavity is chosen to have the same frequency as the main linac (11.9942 GHz) as a compromise between size, phase stability requirements and beam loading. It is proposed to use a HE11 mode travelling wave structure as the CLIC crab cavity in order to minimise beam loading and mode separation. The position of the crab cavity close to the final focus enhances the effect of transverse wake-fields so effective wake-field damping is required. A damped detuned structure is proposed to suppress and de-cohere the wake-field hence reducing their effect. Design considerations for the CLIC crab cavity will be discussed as well as the proposed high power testing of these structures at SLAC.

## INTRODUCTION

A crab cavity is a transverse deflecting dipole cavity used to rotate particle bunches prior to collision where the IP has a finite crossing angle. Crab cavities typically operate using the $TM_{110}$ like hybrid mode of an RF cavity, in which the transverse electric and magnetic fields act together to kick the bunches in the same plane, shown in Figure 1.

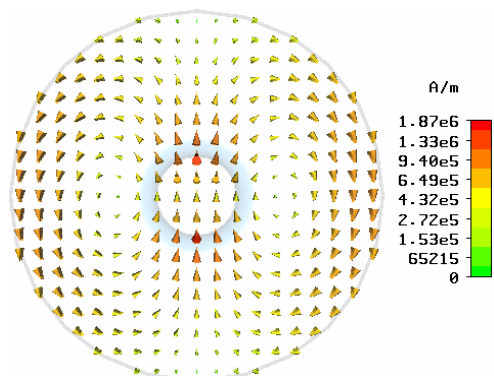

Figure 1: The magnetic field of a TM110-like hybrid dipole mode.

Crab cavities were first proposed by Palmer in 1988 [1] for the rotation of particle bunches to preserve luminosity at the IP for lepton colliders. This concept was successfully demonstrated at KEK-B [2] in 2007 using two single cell deflecting cavities. A crab cavity can also be used to rotate the bunch in a light source for the production of short X-rays [3]. Crab cavities are now proposed at a number of accelerators including ILC [4], LHC [5], APS [3] and CLIC.

In the case of CLIC the 20 mrad crossing angle coupled with the small bunch dimensions would result in a 90 % luminosity reduction if not corrected. A pair of crab cavities placed before the final focus in each beamline is proposed to rotate the bunches prior to collision.

In this paper the RF requirements for a CLIC crab cavity are analysed and the design of an X-band travelling wave dipole cavity is discussed including wakefields, damping and couplers.

## CLIC REQUIREMENTS

The phase stability requirements for CLIC are tighter than for any other crab cavity proposed. The phase stability requirements for a crab cavity are given by equ 1 where S is the luminosity reduction factor, $\sigma_x$ is the horizontal bunch size at the IP, $\theta_c$ is the crossing angle, and $\omega$ is the cavity angular frequency [6].

$$\Delta\varphi < \frac{4\omega\sigma_x \sqrt{\ln(1/S)}}{c\,\theta_c} \qquad (1)$$

as can be seen in this equation that the phase stability tolerances increase with increasing frequency. However due to the need for low wakefields and accurate machining the cavity frequency cannot be set too high. For the CLIC crab cavity it has been initially chosen to operate at the main linac frequency of 11.9942 GHz as a compromise between phase stability and cavity size. For the CLIC beam size of 60 nm the timing stability is 5 fs for a 2 % luminosity loss which is a major challenge to be overcome and will certainly require all cavities to be driven by a single amplifier.

The amplitude tolerance of a crab cavity is set by the luminosity loss associated with beams colliding with crossing angles. The incorrect amplitude on a crab cavity will cause incorrect bunch rotation for the crossing angle and the bunches will collide with a small angle between them. The tolerable amplitude stability is given in equation 2

$$\frac{\Delta V}{V_{cav}} = \frac{2}{\theta_c} \frac{\sigma_x}{\sigma_z} \sqrt{\frac{1}{S^2} - 1} \qquad (2)$$

This leads to an amplitude tolerance of 2.0 % for the CLIC crab cavities which should not prove difficult to achieve.

The displacement of an electron at the IP, $\Delta x$, leading or trailing the bunch centroid by a time $\Delta t$, caused by a crab cavity is given by,

$$\Delta x(\Delta t) = R_{12} \frac{V_{cav}}{E_o} \sin(\omega \Delta t) \qquad (3)$$

where $R_{12}$ is the ratio of the bunch displacement at the IP to the divergence created by the crab cavity. The voltage, $V_{cav}$, required to cancel the crossing angle of a bunch of energy, $E_0$, is given by equation 4,

$$V_{cav} = \frac{cE_0 \theta_c}{2\omega R_{12}} \qquad (4)$$

where $\theta_c$ is the crossing angle, and $\omega$ is the cavity frequency. The crab cavity is positioned at a location with a high $R_{12}$ to reduce the required voltage. The CLIC has a crossing angle of 20 mrad and an $R_{12}$ of 25 m, hence a 11.9942 GHz cavity will require a voltage of 2.39 MV at 3 TeV CoM.

## BEAM-LOADING IN DIPOLE CAVITIES

Dipole modes have a longitudinal electric field that is zero on axis. This means in theory there should be negligible beam-loading in dipole cavities. However in practice the beam does not typically traverse the cavity on axis and often has a small offset. This small offset is often largest in regions with high beta functions such as the location of the crab cavity. The longitudinal electric field of a crab cavity varies linearly with radius for small offsets less than the iris radius, hence the beam-loading will also vary linearly with beam offset. It is unlikely that the beam offset will be constant and is likely to vary considerably train-to-train and hence the beam-loading will be different for every bunch train.

Unlike deflecting mode cavities, the beam induced voltage in a crab cavity will be in-phase with the crabbing mode hence, unless the bunch is significantly early or late, the beam-loading will have a large effect on the crabbing amplitude rather than its phase. The amplitude of the beam-loading, if not removed quickly, could easily alter the crabbing amplitude by more than the 2 % tolerance hence it is necessary to reduce the cavity filling time by using a structure with a high group velocity , and hence a travelling wave structure.

## CELL SHAPE OPTIMISATION

The cell shape was chosen to be a simple iris loaded cavity. The cavity iris radius, iris thickness and cavity radius could be varied to alter the cavity properties and the cavity length was fixed by the phase advance. A parameter sweep for two phase advances ($2\pi/3$ and $5\pi/6$) was performed in Microwave Studio [7] altering the iris radius and thickness and varying the cavity radius to keep the frequency constant at 11.9942 GHz. The peak fields, shunt impedance, Q factor and group velocity were recorded for each parameter set. The iris radius was varied from 2 to 5 mm and the iris thickness from 1 to 8 mm. The simulations used a single cell with periodic boundaries and a mesh of 35 lines per wavelength. A future study will investigate $\pi/2$ structures which are likely to increase the group velocity at the expense of surface fields and R/Q. A standing wave structure was also simulated but is not considered here.

As expected the $2\pi/3$ mode had a higher group velocity than the $5\pi/6$ mode and the velocity decreased with iris thickness and increased with iris radius for iris radii below 4 mm. The group velocity peaks for iris radii between 4 and 5 mm depending on the iris thickness and phase advance and then decreases as the iris radius increases, as shown in Figure 2.

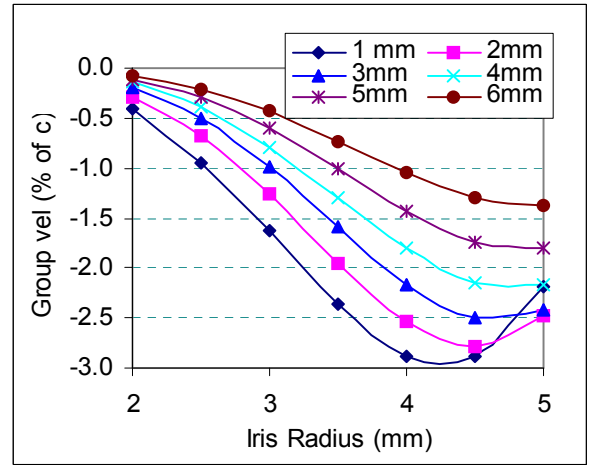

Figure 2: The group velocity of a $2\pi/3$ structure as a function of iris radius, for various values of iris thickness.

Due to the tight phase and amplitude tolerances and the large variation of beam loading train to train in crab cavities it is advantageous to have a large group velocity suggesting a thin iris of between 1 to 3 mm with a radius between 4 and 5 mm and a phase advance of 2π/3. The shunt impedance was also maximised for thin iris'. At iris radii of 4 to 5 mm the 2π/3 mode has very large peak fields on the iris for a 1mm iris thickness so this thickness was also discounted.

Initially a structure was designed at 11.424 GHz to match the SLAC X-band klystrons on which we hope to test our structure. This structure had an iris thickness of 2 mm and an iris radius of 5 mm. This gave the structure a ratio of transverse gradient to peak electric field of 0.3, a group velocity of 3.2% of c (the speed of light in a vacuum) and a transverse R/Q of 58 Ohms.

The azimuthal polarisation of the crab cavity must be well aligned to the crossing angle to avoid spurious vertical rotation, which can cause a loss of luminosity [8]. In order to keep the polarisation of the kick well defined and to aid in the damping of the same order mode (SOM), the vertical polarisation of the crabbing mode, the cavity must be made azimuthally asymmetric. This can be achieved in a number of ways
   1. :Making the cavity elliptical
   2. Asymmetric damping (using waveguides, manifolds or asymmetric chokes)
   3. Polarising rods
   4. Coupling slots between cells

Making the cavities elliptical is likely to increase the machining cost and decrease machining accuracy as the structures could no longer be machined on a lathe, however this will give the best field profile and a large frequency separation. Asymmetric damping will be discussed later in this paper but this option provides good SOM damping and polarisation at the same time. Polarising rods or slots is a simple and easy way of polarising the cavity but could enhance surface fields.

## SINGLE BUNCH WAKEFIELDS

The small bunch size at the CLIC IP will make the beam very sensitive to transverse offsets. As the CLIC crab cavity is proposed to be at X-band, which requires small iris radii, and the cavity is positioned just before the final focus, where the horizontal beta function and hence $R_{12}$ is at its largest, the wakefields could potentially cause a large luminosity loss. Large wakefields would force the cavity to be designed at a lower frequency, hence making the phase stability requirement unfeasibly hard. In order to quantify the short range wakefields the finite difference code ECHO2D was used to calculate the longitudinal and transverse wakefields induced by the CLIC bunch in a dipole cavity with various iris radii. A mesh of 30 lines per sigma was used longitudinally and a bunch length of 45 μm was used. It was found that for a 2 mm iris radius the transverse kick was 2021 V/pC/m and for a 5 mm radius the kick was 335 V/pC/m for a 7 cell cavity, shown in Figure 3.

Assuming the wakefield scales linearly with the number of cells, the luminosity loss for a 30 cell cavity, assuming a bunch charge of 0.6 nC, a beam energy of 0.5 TeV and a horizontal bunch offset of 0.25 mm at the crab cavity, was 2% for a 5mm iris. The offset at the IP due to the wake doubles for a 4mm iris radius so this is likely to be too small an iris to be considered. The longitudinal wakefield is likely to be less problematic and well within tolerable limits. The longitudinal wakefield for a 5 mm iris was calculated to be 250 V/pC for the 7 cell cavity. Assuming 30 cells this is a voltage of 0.6 MV for a 5 mm iris radius.

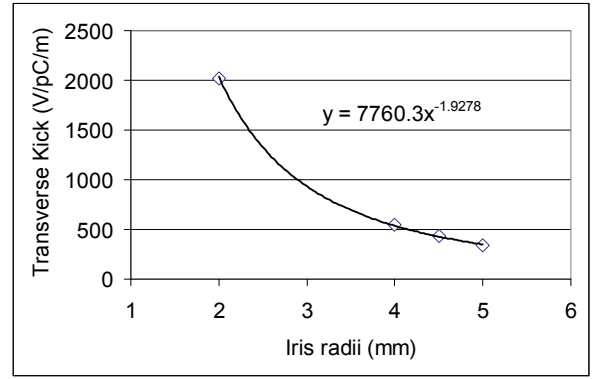

Figure 3: Transverse Kick as a function of iris radius.

## MULTI-BUNCH WAKEFIELDS IN DIPOLE CAVITIES

The multi bunch wakefields are likely to be much more problematic and will require a large amount of damping to reduce the wake substantially, similar to the main linac structures [9]. This task is complicated for the crab cavity by the lower and same order modes (LOM's and SOM's) in addition to the usual higher order modes (HOM's). A study of the HOM's in a single cell with periodic boundaries showed that the 4th dipole passband had a particularly high R/Q. The R/Q of this mode was found to vary substantially with small modifications of the cavity geometry and in future shape optimisations this mode will be considered.

In addition to damping of the SOM the mode may be detuned in order to reduce the wake faster by destructive interference. This is achieved by giving the SOM a different frequency in each cell while keeping the crabbing mode frequency constant. This can be achieved in practice by varying the degree of polarisation in each cell.

Several damping schemes were investigated for the mode damping. In this paper we will discuss choke structure damping, waveguide damping, and iris damping.

The choke structure, figure 4, has a SiC ring in the cavity design to absorb any unwanted mode, with the operating mode shielded by a choke filter. This was simulated in Microwave studio using a single cell with periodic boundaries, and a lossy eigenmode simulation was performed to calculate the external Q factor of the modes. The choke mode was found to provide good damping of the LOM giving Q factors as low as 50, but unless the cavity is polarised azimuthally the SOM will not be damped. Initially inserting polarising rod in the cavity was attempted but this was found to have little effect on the SOM Q. Next having a symmetrical cavity and an asymmetric choke was simulated and this was found to lower the SOM external Q substantially to around 500, however the LOM external Q rose to about the same level.

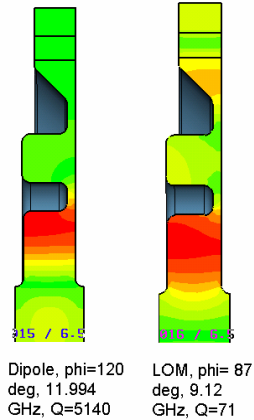

Figure 4: The LOM and crabbing mode in a choke mode structure.

Waveguide and manifold damping has the added advantage of polarising the cavity and providing stronger damping. The waveguide damping was found to damp the LOM to an external Q of around 150 and the SOM Q to ~30. However in order to avoid damping the crabbing mode the waveguides can only be placed perpendicular to the polarisation of the crabbing mode in the vertical plane. In the horizontal plane cut-off waveguide can be used but this may not be as effective at damping the HOM's. The dipole modes in the vertical plane will require the most damping due to the presence of the SOM and the smaller bunch size in that plane.

Neither of the methods above will damp the TE111-like dipole mode which has most of its fields concentrated in the iris. This mode can be effectively be damped by using bi-periodic cavities with lossy ceramics placed in the shorter of the cavities. Iris damping will however, as well as difficulties in manufacturing, reduce the structure group velocity as it will require thicker iris', hence is not an optimal damping method for the CLIC crab cavity. As the TE111-like dipole mode has a very high group velocity and a low R/Q it is probably acceptable to damp this mode with couplers in the beampipe.

## STRUCTURE FOR HIGH POWER TESTS

To date, there has been a huge amount of effort and resources focussed on high gradient tests of accelerating mode cavities, however there is relatively little known about maximum gradients in dipole cavities other than what can be inferred from the monopole measurements and a number of dipole measurements at 3 GHz performed by CERN [10,11]. In order to obtain more information on the high gradient performance of these structures at X-band it is proposed to test a 7 cell travelling wave deflecting cavity at SLAC. This structure will have 5 crab cavity mid-cells and special $TE_{111}$ mode matching cells, shown in Fig 5, to ensure low fields at the waveguide coupler and in the matching cells themselves.

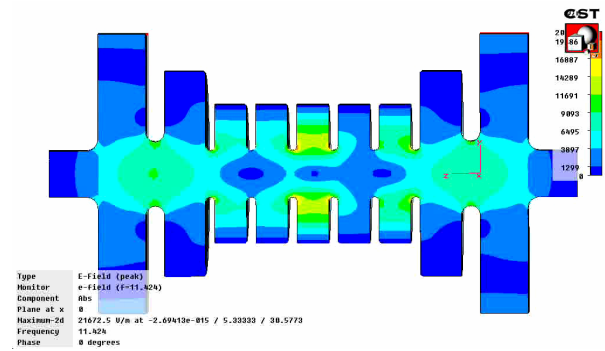

Figure 5: The crab structure designed for High power tests at SLAC.

This structure achieves a peak electric field of 110 MV/m and a peak magnetic field of 350 kA/m for an input power of 20 MW, which corresponds to a transverse kick of 1.62 MV over the 5 cells this gives a transverse gradient of 37 MV/m.

For CLIC and preliminary tests on CTF3 we would like to reduce the overall length of the structure while keeping the transverse voltage high. In order to decrease the length of the structure the matching cells should now have a similar gradient to the mid cells and the waveguides should be coupled directly into the equator of the matching cell. This requires a complex optimisation of the matching cell geometry such that the cavity is matched and has a similar field amplitude in the mid and matching cells. This requires the variation of three parameters to achieve the requirements. It was chosen to optimise the coupling slot width between the waveguide and the matching cell, the matching cell radius and the matching cell iris radius.

## CONCLUSION

Design of a crab cavity for CLIC is underway at the Cockcroft Institute in collaboration with SLAC. This effort draws on a large degree of synergy with the ILC crab cavity developed at the Cockcroft Institute and other deflecting structure development at SLAC.

A study of phase and amplitude variations in the cavity suggests that the tolerances are very tight and require a 'beyond state of the art' LLRF control system.

A study of cavity geometry and its effect on the cavity fields has been performed using Microwave studio. This study has suggested that for our cavity an iris radius between 4-5 mm is optimum with an iris thickness of 2-3 mm based on group velocity and peak fields.

A study of the cavity wakefields show that the single bunch wakes are unlikely to be a problem but the short bunch spacing may cause the multi-bunch wakefields to be an issue. This will require some of the modes to be damped strongly so that the wake is damped significantly before any following bunch arrives. Various methods of damping have been investigated and suggest that waveguide damping in the cells should provide sufficient damping in the vertical plane, which is the most sensitive.

## ACKNOWLEDGEMENTS

The authors would like to thank Daniel Schulte at CERN and Andrei Seryi at SLAC for their help in this work. This work is supported by STFC and FP7 EUCARD.